\documentclass[a4paper,12pt]{article} \tolerance=200
\usepackage{amsfonts} 
\title{Noncommutative effective theory of vortices in a 
complex scalar field}
\author{C.~D.~Fosco$^a$~\footnote{Electronic address:
    fosco@cab.cnea.gov.ar} and A.~L{\'o}pez$^a$~\footnote{Electronic
    address: lopezana@cab.cnea.gov.ar}
  \\ \\
  {\normalsize\it $^b$Centro At{\'o}mico Bariloche - Instituto Balseiro,}\\
  {\normalsize\it Comisi{\'o}n Nacional de Energ{\'\i}a At{\'o}mica}\\
  {\normalsize\it 8400 Bariloche, Argentina.}}  \date{}
\begin{document}
\maketitle
\begin{abstract}
\noindent We derive a noncommutative theory description for vortex
configurations in a complex field in $2+1$ dimensions.  We interpret
the Magnus force in terms of the noncommutativity, and obtain some
results for the quantum dynamics of the system of vortices in that
context.
\end{abstract}
\newpage 
Noncommutative field theories have recently been the subject of
intense research, mostly because of their relevance to the study of
some situations arising in string theory, where an antisymmetric
tensor field coupled to the world sheet develops a non-zero vacuum
expectation value~\cite{DN,castellani}. On the other hand, it has been
realized that the noncommutative geometry setting may also be useful
in the description of the quantum Hall
effect~\cite{susskind,AP,KS,MaSt}, where the verification of
noncommutative geometry effects should be more accessible from the
experimental point of view than in the previous example.

The emergence of noncommutativity is usually understood as a
consequence of the presence of a strong magnetic field, a role which
in the string theory context is played by the antisymmetric tensor
field, and by the real magnetic field in the quantum Hall effect. It
is our purpose in this letter to emphasize that a noncommutative
theory may also be a good approximation to the description of vortices
in a planar (i.e., two spatial dimensions) complex field.  Here, a
dual description shall be introduced, such that the `magnetic field'
will actually be due to the non-zero vacuum density of the field.

Our starting point is the action for a self-interacting
nonrelativistic complex field $\phi$:
\begin{equation}\label{eq:defs} 
S(\phi^*,\phi) \;=\; \int dt d^2x
\left[
\frac{i\hbar}{2} (\phi^* \partial_0 \phi - \partial_0 \phi^*\phi) 
- \frac{\hbar^2}{2 m}\partial_j\phi^* \partial_j \phi - V(\phi^*\phi) 
\right]
\end{equation}
where $V(\phi^* \phi) = \frac{\lambda}{2}(\phi^* \phi - \mu)^2$, with $\mu > 0$, is a
potential which favors the emergence of a non zero expectation value
for the charge density of the field: $\rho\,=\,\phi^*\phi$. To consider vortex
configurations, we use the technique and notation that have been used
for a similar system in~\cite{Zhang,MS}. What follows is an adapted
version of that technique to the case at hand.

We thus introduce a more convenient
parametrization of the scalar field:
\begin{equation}\label{eq:phipar}
\phi(x) \;=\; \sqrt{\rho (x)} \, e^{i \theta (x)} \, {\tilde \phi}(x)
\end{equation}
in terms of the density; a field $\theta$ which is the regular part of
the phase of $\phi$; and ${\tilde \phi}$, which accounts for the
singular part of the configuration. The singular part is, of course,
constrained to have modulus equal to one.

Introducing the parametrization (\ref{eq:phipar}) into
(\ref{eq:defs}), we find an equivalent expression for the action,
namely:
$$
S=\int dt d^2x \left[- \hbar \rho \partial_0 \theta + i \hbar \rho {\tilde \phi}^* \partial_0 {\tilde \phi} \right.
$$
\begin{equation}\label{eq:sequiv}
\left. - \frac{\hbar^2\rho}{2 m} (\partial_j \theta - i {\tilde \phi}^* \partial_j {\tilde \phi})^2 -
\frac{\hbar^2}{8 m \rho} \partial_j \rho \partial_j \rho - V(\rho) \right].
\end{equation}
To linearize the third term in (\ref{eq:sequiv}), we introduce an 
auxiliary
vector field ${\vec J}$, so that
$$
S\;=\; \int dt d^2x \left[- \hbar \rho \partial_0 \theta + i \hbar
\rho {\tilde \phi}^* \partial_0 {\tilde \phi} - \hbar J_k (\partial_k
\theta - i {\tilde \phi}^* \partial_k {\tilde \phi}) \right.
$$
\begin{equation}\label{eq:sequiv1}
\left. + \frac{m}{2 \rho} J_k J_k \,-\,\frac{\hbar^2}{8 m \rho}
\partial_j \rho \partial_j \rho - V(\rho) \right]\;.
\end{equation}
The regular phase $\theta$ becomes now a Lagrange multiplier field,
which imposes the linear constraint $\partial_0 \rho + \partial_k J_k
= 0$, i.e., a continuity equation mixing the density associated to the
field $\phi$ with a `current' defined by the auxiliary field ${\vec
J}$. This constraint may be solved by introducing a field $b_\mu$,
such that $\rho=\epsilon_{jk}\partial_j b_k$ and
$J_k=\epsilon_{kl}(\partial_l b_0 - \partial_0 b_l)$. This will not,
of course, determine $b_\mu$ completely, since the equations that give
$\rho$ and $J_k$ are invariant under the `gauge transformations':
$b_\mu \to b_\mu + \partial_\mu \alpha$.  This freedom should, and
indeed will, be taken into account for the derivation of the effective
theory, by adopting a gauge fixing condition whenever required (i.e.,
when inverting operators depending on the quadratic form for $b_\mu$).

After some elementary algebra, the action $S$ may be written in the
form:
\begin{equation}\label{eq:sequiv2}
S\;=\; \int dt d^2x \, \left[ -h b_0 {\tilde J}_0 - h b_k {\tilde J}_k
+ \frac{m}{2 \rho_b} {\vec J}^2_b - \frac{\hbar^2}{8 m \rho_b}
\partial_j\rho_b \partial_j \rho_b \,-\, V(\rho_b) \right] \;,
\end{equation}
where ${\tilde J}_\mu$ is a `topological' vortex current, with
components defined by
\begin{equation}\label{eq:defjtop}
{\tilde J}_0 \,=\, \frac{1}{2\pi i} \, \epsilon_{kl} \partial_k
({\tilde \phi}^* \partial_l {\tilde \phi} ) \;\;\; {\tilde J}_k
\,=\,\frac{1}{2\pi i} \,\epsilon_{kl} [\partial_l ({\tilde \phi}^*
\partial_0 {\tilde \phi} ) - \partial_0 ({\tilde \phi}^* \partial_l
{\tilde \phi})]
\end{equation}
and the notation $\rho_b$ and ${\vec J}_b$ has been used to emphasize
the fact that those fields are determined by $b_\mu$, since $\theta$
has been integrated out, and the corresponding constraint completely
solved.

Of course, the topological current can only be non-vanishing when the
\mbox{$\phi$-field} configuration has singularities. In two spatial
dimensions, they can only correspond to isolated points ${\vec
x}^{(\alpha)}$, around which the phase winds up an integer number of
times, $q_\alpha$~\footnote{The sign of $q_\alpha$ is defined as
positive when the phase winding is counterclockwise.}.  Thus, an $N$
vortex configuration carries a vortex density and current which may be
written as follows:
\begin{equation}\label{eq:vcur}
{\tilde J}_0(t,x) \;=\; \sum_{\alpha =1}^N q_\alpha \delta [{\vec
x}-{\vec x}^{(\alpha)}(t)]\;,\;\; {\tilde J}_k(t,x) \;=\; \sum_{\alpha
=1}^N q_\alpha \, \dot{x}_k^{(\alpha)} \delta [{\vec x}-{\vec
x}^{(\alpha)}(t)] \;,
\end{equation}
where ${\vec x}^{(\alpha)}(t)$ denotes the actual trajectory of the
vortex labeled by the index $\alpha$.  The quantum dynamics of the
vortices corresponding to this current shall be derived by considering
the functional integral representation of the vacuum transition
amplitude, and integrating out the auxiliary field $b_\mu$. Namely,
the action describing the effective dynamics of the vortices,
$S_{eff}$, will be expressed as:
\begin{equation}\label{eq:defseff}
\exp\{\frac{i}{\hbar} S_{eff}[{\tilde J}_\mu]\}\;=\; \int [{\mathcal
D}b_\mu] \; \exp \{\frac{i}{\hbar} S[b_\mu; {\tilde J}_\mu]\} \;,
\end{equation}
where the symbol $[{\mathcal D}b_\mu]$ has been used to denote the
$b_\mu$ field integration measure, including gauge fixing artifacts
(i.e., Faddeev-Popov factors). They will be neglected in the following
discussion, since the theory is Abelian and hence they factor out for 
the calculations we are interested in.  $S_{eff}$ cannot be evaluated
exactly, because of the presence of non-quadratic terms in the action. 
We may, however, obtain an approximation to $S_{eff}$ by integrating 
out the quadratic fluctuations around an homogeneous vacuum 
configuration for $b_\mu$. Taking into account the special form we 
have assumed for the potential $V$, we see that the homogeneous 
vacuum corresponds to a constant density \mbox{$\rho_b \equiv \rho_0 = \mu$}
which may be solved for $b_j$, by adopting the symmetric gauge:
\mbox{$b_j = -\frac{\rho_0}{2} \epsilon_{jk} x_k$}.

Assuming first the vacuum to be isotropic, the current ${\vec J}_b$
must of course vanish. Moreover, as ${\vec b}$ is time-independent,
the spatial components of the current are: $J_k = \epsilon_{kl}
\partial_k b_0$, and any constant $b_0$ value becomes compatible with
this vacuum configuration.  This constant will then have to be
integrated out. Denoting by $S^{(0)}_{eff}$ the result of evaluating
the action on this vacuum configuration, we see that
\begin{equation}\label{eq:s0eff}
S^{(0)}_{eff} \;=\; \int dt \, \frac{h \rho_0}{2} \epsilon_{kl} x_l {\tilde J}_k  \;,
\end{equation}
plus the constraint:
\begin{equation}\label{eq:constraint}
Q \,\equiv\, \int d^2x \, J_0 \,=\, 0
\end{equation}
which follows from the integration of the constant value
$b_0$~\footnote{We are assuming the gauge invariance to hold even at
spatial infinity, i.e., the gauge group is not reduced to the
identity at infinity. This is the reason why the net charge has
to be zero. This, on the other hand, guarantees the finiteness of
the static energy.}.

We shall now include fluctuations on top of this configuration.  To
that end, we consider small variations $\delta\rho$ and $\delta{\vec
J}$, which may be written in terms of $\delta b_\mu$ by adopting a
gauge. We then expand the action $S$ up to the quadratic order in
$\delta b_\mu$, and integrate out $\delta b_\mu$. In this Gaussian
approximation, and ignoring ${\tilde J}$-independent terms, we see
that the effect of these fluctuations amounts to adding to
$S^{(0)}_{eff}$ a contribution $S^{(1)}_{eff}$, given explicitly by
$$
S^{(1)}_{eff}\;=\; \frac{h^2\rho_0}{2 m} \int dt d^2x d^2x'\, {\tilde J}_0(t,x) \Delta^{-1}(x,x')
{\tilde J}_0(t,x')
$$
\begin{equation}\label{eq:s1eff}
+ \frac{h^2}{2} \int dt dt'\, d^2x d^2x'\, {\tilde J}_k (t,x)
K_{kl}(t-t',x-x') {\tilde J}_l (t',x') \;,
\end{equation}
where
\begin{equation}\label{eq:defk}
 K_{kl}(t,x)= \delta_{kl} \int\frac{d\omega d^2p}{(2 \pi)^3}
\frac{e^{-i\omega t + i {\vec p}\cdot{\vec x}}}{(\frac{\hbar^2 p^2}{4
m \rho_0} + \lambda) p^2 - \frac{m}{\rho_0} \omega^2}
\end{equation}
is an operator containing nonlocalities in time~\footnote{We are
omitting the (implicit) $i \epsilon$ to avoid the pole.}. To
obtain a local theory, we assume that $\lambda$ is sufficiently large,
so that only the first term in (\ref{eq:s1eff}) (which is
$\lambda$-independent) remains.  This yields a (time local) expression
for $S^{(1)}_{eff}$:
\begin{equation}\label{eq:s2eff}
S^{(1)}_{eff}\;=\; \frac{h^2\rho_0}{2 m} \int dt d^2x d^2x'\,{\tilde
J}_0(t,x) \Delta^{-1}(x,x') {\tilde J}_0(t,x')
\end{equation}
where the first correction to this expression is of order $\sim
\lambda^{-1}$
\begin{equation}\label{eq:s3eff}
{\mathcal O}(\lambda^{-1})\;=\; - \frac{h^2}{2 \lambda} \int dt d^2x  d^2x'\, {\tilde J}_k (t,x) \Delta^{-1}(x-x')
{\tilde J}_k (t,x') \;.
\end{equation}
This large-$\lambda$ approximation may be justified if the condition
\mbox{$\frac{m v^2}{\rho_0 \lambda} << 1$} (where $v$ denotes the
typical velocity of the vortices) holds.

Putting together the first two leading terms contributing to the
effective action, we get a tractable approximation to the effective
action:
\begin{equation}\label{eq:seff}
S_{eff}=\int dt \left[ \frac{h \rho_0}{2}\epsilon_{kl} x_l {\tilde J}_k +\frac{h^2 \rho_0}{2 m}
\int d^2x d^2x'{\tilde J}_0(t,x) \Delta^{-1}(x,x') {\tilde J}_0(t,x') \right] \,,
\end{equation}
which should be supplemented by the constraint that imposes the
vanishing of the total vortex charge.  Notice that even for finite $\lambda$
the second term in (\ref{eq:s1eff}) will not modify the phase space
structure of the theory described by (\ref{eq:seff}).  Indeed, its
expansion is non-local in time, and there are no quadratic term in the
time derivatives.

We will use this expression as a starting point for the noncommutative
description of the vortex dynamics in this model.  Let us first
consider a `first quantized' language: introducing in (\ref{eq:seff})
expression (\ref{eq:vcur}) for the vortex current, we see that the
effective action for $N$ vortices becomes
\begin{equation}\label{eq:fq}
S_{eff}\;=\; \int dt \left[ \frac{h \rho_0}{2} \sum_{\alpha = 1}^N
q_\alpha \, \epsilon_{j k} \dot{x}^{(\alpha)}_j x^{(\alpha)}_k +
\frac{h^2\rho_0}{4\pi m}\, \sum_{\alpha,\beta=1}^N q_\alpha q_\beta \,
\ln |\frac{x^{(\alpha)} - x^{(\beta)}}{\xi}| \right]
\end{equation}
where we have written explicitly the inverse of $\Delta$ in the
coordinate representation, and the global neutrality condition is
taken into account by assuming that $\sum_{\alpha=1}^N q_\alpha =0$. The parameter
$\xi$ can be regarded as a minimum length, related to the size of the
vortex core, as in reference~\cite{HW}. 
This action has a very particular structure since the kinetic term
is linear in time derivatives. This means that the piece without time
derivatives is (minus) the Hamiltonian; namely, the effective {\em
first order\/} Lagrangian is:
\begin{equation}\label{eq:fq1}
{\mathcal L}_{eff}\;=\; \frac{h \rho_0}{2} \sum_{\alpha = 1}^N
q_\alpha \, \epsilon_{kl} \dot{x}^{(\alpha)}_k x^{(\alpha)}_l -
{\mathcal H}
\end{equation}
with
\begin{equation}
{\mathcal H}\;=\; - \frac{h^2\rho_0}{4\pi m}\, \sum_{\alpha,\beta=1}^N q_\alpha q_\beta \, 
\ln |\frac{x^{(\alpha)} - x^{(\beta)}}{\xi}| \;,
\end{equation}
which coincides with the one of  ref \cite{fetter,HW}, where it was obtained using a hydrodynamic approach  for a system of slightly deformed rectilinear vortices in an incompressible fluid.

The special form of the term containing the time derivatives is
responsible for the noncommutativity, since the canonical commutators
following from (\ref{eq:fq1}) are:
\begin{equation}\label{eq:canco}
[x^{(\alpha)}_k , x^{(\beta)}_l] \;=\; \frac{i}{2 \pi \rho_0 q^{(\alpha)}}\, \delta_{\alpha \beta}\, \epsilon_{kl} \;\equiv\;
\frac{i}{q^{\alpha}} \, \theta \, \delta_{\alpha \beta}\,
\epsilon_{kl}
\end{equation}
(no sum over $\alpha$). Contrary to what happens when it is due to a real
magnetic field, here the noncommutativity parameter $\theta=(2\pi \rho_0)^{-1}$
is purely classical, i.e., there is no $\hbar$ factor~\footnote{Of course,
  the commutator between $x_k$ and its canonical conjugate is
  proportional to $\hbar$. The presence of an $\hbar$ factor in the
  Lagrangian, however, cancels out the $\hbar$ factor in the commutator
  between coordinates.}. This is also consistent with dimensional
analysis, since $q$ is dimensionless and $\rho_0^{-1}$ has the dimensions
of an area. By contrast, in the magnetic field induced
noncommutativity, the analogous relation yields a $\theta$ proportional to
$l^2$, where $l$ denotes the cyclotron length, which is proportional
to $\hbar^{\frac{1}{2}}$.

The Hamiltonian ${\mathcal H}$ is the one of a two-dimensional neutral
Coulomb gas, with the neutrality condition guaranteeing the finiteness
of the energy.  It should be noted that we are here considering a
sector corresponding to a (fixed) number ($N$) of vortices. Had we
wanted to compare sectors with different numbers of singularities, the
constant terms we have neglected would have been relevant (i.e., to
determine the chemical potential of the system). The partition
function for the classical system of vortices at finite temperature is
then proportional to the one of a (globally neutral) two dimensional
Coulomb gas. It is noteworthy that this partition function is not
exactly identical to the one of a standard, classical $2+1$
dimensional Coulomb gas composed of dynamical charges, since that
theory would have a phase space twice as large, because the canonical
momenta would be independent from the coordinates, at least assuming
normal quadratic kinetic energy terms. The Gaussian integral over the
canonical momenta would then, as usual, yield a decoupled factor which
modifies the total entropy, as well as the zero of the free energy.

It is, at this level, already possible to make contact with some
noncommutative geometry results. In particular, the interpretation of
noncommutative field theory as describing elementary dipoles~\cite{BS}
is found by looking at the simplest case: $N=2$, i.e., a 
vortex-antivortex pair
\begin{equation}\label{eq:fq2}
{\mathcal L}_{eff}\;=\; \frac{h \rho_0 q}{2} \epsilon_{kl}
[\dot{x}^{(1)}_k x^{(1)}_l - \dot{x}^{(2)}_k x^{(2)}_l]
\,-\,\frac{h^2\rho_0 q^2}{2\pi m}\, 
\ln | \frac{x^{(1)} - x^{(2)}}{\xi} | \;,
\end{equation}
where $x^{(1)}$ and $x^{(2)}$ denote the vortex ($q_1 = q$) and
antivortex ($q_2=-q$) coordinates, respectively. Introducing the
relative ($r$) and center of mass ($R$) coordinates, we see that
\begin{equation}
{\mathcal L}_{eff}\;=\; h \rho_0 q \, \epsilon_{kl} \dot{r}_k R_l
\,-\,\frac{h^2\rho_0 q^2}{2\pi m}\, \ln |\frac{r}{\xi}| \;,
\end{equation}
where $r_k=x^{(1)}_k - x^{(2)}_k$ and $R_k = \frac{x^{(1)}_k +
x^{(2)}_k}{2}$.  This form of the Lagrangian makes it explicit the
fact that the center of mass and relative coordinates of the `dipole'
built from the vortex-antivortex pair become conjugate canonical
variables. We also note that the interaction potential between vortex 
and antivortex is confining, so it makes sense to attempt a description
in terms of the dipole as a single entity, with the noncommutativity 
taking into account part of the vortex `internal structure'.

For more tractable forms of the interaction potential, like
a quadratic one, the system is equivalent to a single particle with
the coordinates of the center of mass $R_k$ and a mass determined by
the parameter of the harmonic interaction potential~\cite{BS}.
In this interpretation, the uncertainty relations
\begin{equation}\label{eq:uncert}
\Delta r_1 \, \Delta R_2 \;\geq\; \theta \;\;\;\;\; \Delta r_2 \, \Delta R_1 \;\geq\; \theta \;,
\end{equation}
relate the `size' of the dipoles to the uncertainty of the center of
mass coordinates.  The noncommutative description naturally arises
when considering the quantum version of the theory defined by the
effective action (\ref{eq:seff}).

When a (globally neutral) configuration of the system is such that it
can be approximately described as a set of condensed vortex-antivortex
pairs, the dipole interpretation may also be introduced. Let us assume
that $N$ is even, $N= 2 M$, with $M$ vortices of charge $q=+1$ and $M$
with charge $q=-1$. Introducing coordinates $x^{(\alpha)}$ for the
vortices, and $y^{(\alpha)}$ for the antivortices, we first arrange the
different terms in the Lagrangian in a convenient way:
$$
{\mathcal L}_{eff}\;=\; \frac{h \rho_0 q}{2} \sum_{\alpha = 1}^M \epsilon_{kl} [\dot{x}^{(\alpha)}_k x^{(\alpha)}_l - \dot{y}^{(\alpha)}_k y^{(\alpha)}_l]
\,+\, 
\frac{h^2\rho_0 q^2}{4\pi m}\,\sum_{\alpha,\beta=1}^M \, \ln |\frac{x^{(\alpha)} - x^{(\beta)}}{\xi}| 
$$
$$
+ \frac{h^2\rho_0 q^2}{4\pi m}\,\sum_{\alpha,\beta=1}^M \, \ln |\frac{y^{(\alpha)} - y^{(\beta)}}{\xi}| 
$$
\begin{equation}\label{eq:fq4}
- \frac{h^2\rho_0 q^2}{2\pi m}\,\sum_{\alpha,\beta=1}^M \, \ln |\frac{x^{(\alpha)} - y^{(\beta)}}{\xi}| \;.
\end{equation}
Assuming that the situation is such that the system has condensed into
vortex-antivortex pairs with $x^{(\alpha)}$ paired to $y^{(\alpha)}$, $\forall \alpha$, it
is convenient to introduce the new set of coordinates:~\mbox{${\vec
    r}^{(\alpha)}={\vec x}^{(\alpha)} - {\vec y}^{(\alpha)}$} and {${\vec
    R}^{(\alpha)}=\frac{{\vec x}^{(\alpha)} + {\vec y}^{(\alpha)}}{2}$}. If the
dipoles have condensed, the relative distance $|r^{(\alpha)}|$ should be
negligible in comparison with the distance between the center of
masses of the dipoles. Using a two-dimensional multipole expansion
for the interaction potentials, we have of course
\begin{equation}\label{eq:fq5}
{\mathcal L}_{eff}\;\sim\; h \rho_0 q \sum_{\alpha = 1}^M \epsilon_{kl} \, \dot{r}^{(\alpha)}_k  R^{(\alpha)}_l 
\,-\, \frac{h^2\rho_0 q^2}{2 \pi m}\,\sum_{\alpha=1}^M \, \ln |\frac{r^{(\alpha)}}{\xi}|\;,
\end{equation}
as the leading contribution. The next to leading term introduces a
dipole-dipole interaction potential $V_{\alpha,\beta}$, such that for each pair
$\alpha,\beta$ of dipoles it is given by:
\begin{equation}
V_{\alpha\beta}\;=\; \frac{h^2\rho_0 q^2}{2 \pi m}\, r_j^{(\alpha)} \Omega^{(\alpha\beta)}_{jk} r_k^{(\beta)}
\end{equation}
where
\begin{equation}
\Omega^{(\alpha\beta)}_{jk}\,=\, \frac{2 R^{(\alpha\beta)}_j  R^{(\alpha\beta)}_k - (R^{(\alpha\beta)})^2 \delta_{jk} }{(R^{(\alpha\beta)})^4}\;\;\;,
\;\;\; R^{(\alpha\beta)} \equiv R^{(\alpha)}-R^{(\beta)}\;. 
\end{equation}

We are interested in the quantum theory corresponding to this system.
The convenience of having a `field' description for the many particle
case, should be evident. Indeed, it would be desirable to be able to
define the quantum dynamics in terms of a field corresponding to the
density $J_0$, a function of the coordinates. However, since the
coordinates $x_1$ and $x_2$ of a vortex are conjugate variables, we
see that the density has to be understood as a field defined on a
noncommutative space. The situation is entirely analogous to the
deformation quantization procedure \cite{Zachos} where one describes
the dynamics in terms of the Wigner function, which defines a density
on phase space. In the case at hand, the density will also be a
function of the phase space coordinates $x_1$ and $x_2$, and the
Hamiltonian $H_{eff}$ may be expressed in terms of the Moyal product
$\star$ of the corresponding Weyl symbol:
\begin{equation}
A(x_1,x_2) \star B(x_1,x_2) \,=\, A (x_1,x_2)\, e^{i \frac{\theta}{2} 
({\overleftarrow \partial_{x_1}} {\overrightarrow \partial_{x_2}} 
- {\overleftarrow \partial_{x_2}} {\overrightarrow \partial_{x_1}})}B(x_1,x_2)\;.
\end{equation}
Using the notation $\rho$ for the Weyl symbol of the density, we see that
the Hamiltonian $H$, as obtained from (\ref{eq:seff}), should be
\begin{equation}\label{eq:hnc}
H_{eff}\;=\; \frac{h^2 \rho_0}{2 m} \int d^2x d^2x'\, \rho(t,x)\star \Delta^{-1}(x,x') \star \rho(t,x') \,,
\end{equation}
where, of course, one of the stars may be deleted. Equation
(\ref{eq:hnc}) may be understood as the mean energy of a configuration
defined by the density $\rho$; static solutions that are extrema of
(\ref{eq:hnc}) should be possible collective states of the system. The
equation for those extrema is of course an eigensystem:
\begin{equation}
\frac{1}{4\pi}\int\, d^2y\,  \ln (\frac{|x-y|}{\xi}) \star \rho (y) \;=\; E \, \rho (x) 
\end{equation}  
or, replacing the $*$ by the equivalent shift in the corresponding
coordinate:
\begin{equation}
\frac{1}{4\pi}\int\, d^2y\,  \ln (
\xi^{-1}\, \sqrt{(x_1-i\frac{\theta}{2} D_{y_2})^2 + (x_2+i\frac{\theta}{2} D_{y_1})^2})\; \rho (y) 
\;=\; E \, \rho (x) 
\end{equation}
where $D_{y_1} = \partial_{y_1} + \frac{2 i}{\theta}y_2$ and $D_{y_2} = \partial_{y_2} -
\frac{2 i}{\theta} y_1$ have the form of covariant derivatives for a
constant uniform `magnetic field' $B = \frac{4}{\theta}$.  The presence of
this covariant derivative is indeed a manifestation of the Magnus
force at the purely quantum level, since when considering simplified
versions of the interaction potential (like quadratic ones), it
clearly induces a dynamical behavior similar to the one of particles
in a magnetic field. Of course, the Magnus force is evident at the
semiclassical level just from the particular form of the kinetic term
in (\ref{eq:seff}) \cite{MS}.

Let us, for the sake of completeness, also present the result for the
case of the non-isotropic vacuum with a non-vanishing constant value
$J^0_k$ for the auxiliary field $J_k$. Keeping all the terms which
survive for this configuration, the quadratic approximation yields now
a correction $S^{(1)}_{eff}$, which is explicitly given by
$$
S^{(1)}_{eff}\;=\;  S^{(1)}_{eff}|_{J=0} \,+\, 
\frac{h^2}{2} \int dtd^2x \, dt'd^2x'\,
({\tilde J}_k(t,x) - \frac{J^0_k}{\rho_0} {\tilde J}_0(t,x))
$$
\begin{equation}\label{eq:s4eff}
M^{-1}_{kl}(t,x;t',x')\,({\tilde J}_l(t,x) - \frac{J^0_k}{\rho_0} {\tilde J}_0(t,x))
\end{equation}
where $M_{kl}$ is non-local in time and depends on the (constant)
value of $J_0$.  In momentum space, ${\tilde M}^{-1}$, the Fourier
transform of $M^{-1}$, may be written as
$$
({\tilde M}^{-1})_{kl} (p) \;=\; - \left\{ [ \frac{m}{\rho_0} \omega^2 -
  (\frac{\hbar^2 p^2}{4 m \rho_0} + \lambda) p^2 + \frac {2 m \omega}{\rho_0^2} {\vec J}^0
  \cdot {\vec p} ]^{-1} \; e_{kl} \right.
$$
\begin{equation}
\left. +\; [ \frac{m}{\rho_0} \omega^2 + (\frac{m}{\rho_0^3} (J^0)^2 - \frac{\hbar^2 p^2}{4 m \rho_0} - \lambda) p^2 
+ \frac {2 m \omega}{\rho_0^2} {\vec J}^0 \cdot {\vec p} ]^{-1} \; f_{kl} \right\}
\end{equation}
where $e_{kl} \equiv e_k e_l$, $e_k$: unit vector in the direction of ${\vec J}^0$, and 
$f_{kl} \equiv \delta_{kl} - e_{kl}$.

In the same limit were we found an effective theory local in time
(\ref{eq:seff}) for the ${\vec J}^0 = 0$ case, we find here for the
correction $S_1$:
$$
S^{(1)}_{eff} \;\sim\;\frac{h^2\rho_0}{2 m} \int dt d^2x d^2x'\,{\tilde
  J}_0(t,x) \Delta^{-1}(x,x') {\tilde J}_0(t,x')
$$
\begin{equation}\label{eq:seffj}
- \frac{h^2}{2 \lambda} \int dt d^2x  d^2x'\, ({\tilde J}_k - \frac{J^0_k}{\rho_0} {\tilde J}_0) (t,x) \Delta^{-1}(x-x')
({\tilde J}_k - \frac{J^0_k}{\rho_0} {\tilde J}_0) (t,x')\;.
\end{equation}
This means that the correction to the previous case is still of order
$\lambda^{-1}$, and amount to modifying the vortex current by the
subtraction of a constant current $\frac{J^0_k}{\rho_0} {\tilde J}_0$.

It is interesting to compare the previous results with the ones that
follow by starting from a relativistic model, since there are
similarities but also very important differences between the two
cases. To that end, we consider now the derivation of the effective
theory for vortices in the relativistic scalar field case.  The action
is now assumed to be
\begin{equation}\label{eq:defsr}
S \,=\, \int d^3x \, \left[ \partial_\mu \phi^* \partial_\mu \phi - V(\phi^* \phi) \right]
\end{equation} 
where $V$ is assumed to have the same structure as in the
non-relativistic case, equation~(\ref{eq:defs}). Of course, now the
parameters in that potential have a different physical meaning, but we
keep the same notation for the sake of simplicity.

Introducing in (\ref{eq:defsr}) the same decomposition we have used in
the non-relativistic case, we see that the action $S$ becomes:
\begin{equation}\label{eq:sr1}
S \,=\, \int d^3x \left[ \rho (\partial \theta + i {\tilde \phi}^*
\partial {\tilde \phi})^2   + \frac{1}{4 \rho} (\partial \rho )^2
- V(\rho) \right]
\end{equation}
Again, an auxiliary field is employed to linearize the action. In this
case, we need a vector field $\xi$ so that:
\begin{equation}\label{eq:sr2}
S \,=\, \int d^3x \left[ \xi_\mu (\partial^\mu \theta + i {\tilde \phi}^*
\partial^\mu {\tilde \phi}) - \frac{1}{4 \rho} \xi_\mu \xi^\mu 
+\frac{1}{4\rho} (\partial \rho )^2  - V(\rho) \right] \;.
\end{equation}
Next, the pure gradient $\partial_\mu \theta$ can be replaced by a 
vector field $\theta_\mu \equiv \partial_\mu \theta$ which is, of course,
constrained to verify the zero curl constraint: $\epsilon^{\mu\nu\lambda}
\partial_\nu \theta_\lambda = 0$. This constraint may be enforced by adding
to the action a new term $S_\delta$:
\begin{equation}\label{eq:sdelta}
S_\delta \,=\, \int d^3x \, A_\mu \epsilon^{\mu\nu\lambda} \partial_\nu \theta_\lambda \;,
\end{equation} 
where $A_\mu$ is a Lagrange multiplier field.

This action is assumed to be used for the functional quantization of
the system, so we may, as usual, perform field redefinitions since
they amount to changes in the (functional) integration variables.
Using in particular the shift
\begin{equation}
\theta_\mu \,\to \, \theta_\mu - i {\tilde \phi}^* \partial_\mu {\tilde \phi}
\end{equation}
we find that
$$
S\,=\, \int d^3 \left[ \xi_\mu \theta^\mu - \frac{1}{4\rho} \xi^2 + \frac{1}{4 \rho} 
(\partial \rho)^2 \right.  
$$
\begin{equation}
\left. - V(\rho) + A_\mu \epsilon^{\mu\nu\lambda} \partial_\nu \theta_\lambda + 2 \pi  
A^\mu {\tilde J}_\mu \right]
\end{equation}
where ${\tilde J}_\mu$ denotes the topological current
\begin{equation}
{\tilde J}_\mu \,=\, \frac{1}{2\pi i} \, \epsilon_{\mu\nu\lambda} \partial^\nu 
({\tilde \phi}^* \partial^\lambda {\tilde \phi}) \;.
\end{equation}
Finally, we integrate out the vector field $\theta_\mu$, obtaining,
\begin{equation}
S\,=\, \int d^3x \left[ -\frac{1}{4} (\frac{1}{2\rho}) F_{\mu\nu}(A) F^{\mu\nu}(A) 
+ 2 \pi A^\mu {\tilde J}_\mu +\frac{1}{4 \rho} (\partial \rho)^2 - V(\rho) \right]
\end{equation}
where $F_{\mu\nu}(A) = \partial_\mu A_\nu - \partial_\nu A_\mu$.  This expression is useful in
order to make a comparison with the already considered
non-relativistic case. First we note that, depending on the form
chosen for the potential, one still has the possibility of having a
vacuum where $\rho$ (now a relativistic scalar) has a non-zero uniform
value.  This, however, does not necessarily mean that there will be a
corresponding uniform vacuum magnetic field that guarantees the
desired form for the kinetic term of the action, as in the
non-relativistic case. In other words, to get a non-zero vacuum
magnetic field for $A_\mu$, one should assume that there is spontaneous
breaking of Poincare invariance.  If that is the case, then one can
assume a constant $B = \epsilon_{jk} \partial_j A_k$, a constant $\rho = \rho_0$, and look
for the minima of a `potential' $U$, resulting from adding to $V(\rho_0)$
the contribution coming from the $F^2$ term:
\begin{equation}
U(B,\rho_0) \;=\; V(\rho_0) - \frac{1}{4 \rho_0} B^2 \;.
\end{equation}
The solution for the minima of this potential will provide for an
expression for the uniform magnetic field $B_0$ in terms of the
background density $\rho_0$. Thus, the action evaluated in this
configuration will look exactly like (\ref{eq:s0eff}), but
 with $h\rho_0$ replaced by $2\pi B_0$.
 
 We conclude that a noncommutative geometry description may very well
 be appropriate to the description of vortices in a planar system
 defined in terms of a complex scalar field. The dual description allows one
 to precisely identify all the parameters of the noncommutative
 theory, and moreover the dipole interpretation for the natural
 excitations in a noncommutative theory finds here a natural and
 concrete realization. Also, the introduction of corrections to the
 effective description is under control, and even the existence of
 external currents may be taken into account with minor changes in the
 model.


\end{document}